\documentclass[preprint,preprintnumbers,nofootinbib,aps,10pt,twocolumn]{revtex4-1}
\usepackage{amsmath,amssymb,bm,epsfig}
\usepackage{color}
\usepackage{natbib}
\usepackage{hyperref} 
\usepackage{ulem}
\usepackage{graphicx}
\usepackage{xcolor,colortbl}
\usepackage{lipsum}
\RequirePackage{lineno}
\newcommand{\nn}{\nonumber}
\newcommand{\sNN}{\sqrt{s_{\textrm{NN}}}}

\definecolor{Gray}{gray}{0.85}
\newcolumntype{a}{>{\columncolor{Gray}}c}

\def \beq{\begin{equation}}
\def \eeq{\end{equation}}
\def \beqa{\begin{eqnarray}}
\def \eeqa{\end{eqnarray}}

\begin{document}

\title{ Directed flow of light flavor hadrons for Au+Au collisions at $\sqrt{S_{NN}}=$ 7.7 - 200 GeV}

\author{Tribhuban Parida}
\email{tribhubanp18@iiserbpr.ac.in}
\author{Sandeep Chatterjee}
\email{sandeep@iiserbpr.ac.in}

\affiliation{Department of Physical Sciences,\\
Indian Institute of Science Education and Research Berhampur,\\ 
Transit Campus (Govt ITI), Berhampur-760010, Odisha, India}

\begin{abstract}
We have studied the directed flow of light-flavor hadrons for Au + Au collisions at $\sqrt{S_{NN}}=$ 7.7 - 200 GeV. 
The initial condition is taken from a suitable Glauber model which is further evolved within the framework of relativistic 
hydrodynamics. Model calculations of the rapidity-odd directed flow ($v_1$) of identified light-flavor hadrons are 
compared with the available experimental data after suitably calibrating the initial condition to describe the rapidity 
dependence  of charged particle multiplicity and net-proton yield. For reasonable choice of the initial condition, we are 
able to describe the measured rapidity and beam energy dependence of identified hadron $v_{1}$ including the observed 
$v_1$ splitting between baryons and anti-baryons.

\end{abstract}

\maketitle

\section{Introduction}

Heavy-ion collisions in the Beam Energy Scan (BES) program at the Relativistic Heavy
Ion Collider (RHIC) create a QCD medium with varying amount of net baryon in the mid-rapidity 
region~\cite{STAR:2017sal}, which allows one to probe a large region of the QCD phase diagram. It provides a unique opportunity to understand the dynamics of conserved charges in the QCD medium and to study 
the nature of phase transition from the state of Quark Gluon Plasma (QGP) to hadronic medium at finite 
baryon chemical potential. The beam energy dependence of the rapidity-odd directed flow($v_1$) of identified 
particles is proposed to be one of the relevant observable to probe the nature of QCD phase 
transition~\cite{Rischke:1995pe,Rischke:1996nq,Hung:1994eq,Steinheimer:2014pfa,Ivanov:2014ioa,Ivanov:2016spr}.

The directed flow($v_1(y)$) is defined as the first harmonic coefficient in the Fourier series expansion of the azimuthal
distribution of the particles produced relative to the reaction plane $\Psi_{\text{RP}}$. 
\beq
\frac{d^2 N}{dy d\phi} = \frac{dN}{dy} \left( 1 + 2 \sum_{n=0}^{\infty} v_{n}(y) \cos{ \left[ n(\phi-\Psi_{\text{RP}}) \right] } \right)
\eeq
where $y$ and $\phi$ are longitudinal rapidity and azimuthal angle of a produced particle, respectively. Breaking in 
forward-bacward symmetry generates a non-zero $v_1(y)$ which is an odd function of rapidity in symmetric heavy-ion collisions.

STAR collaboration has measured the energy dependence of the slope of directed flow at mid-rapidity for 
various identified particles~\cite{STAR:2014clz,STAR:2017okv}. A minima in the slope of proton $v_1$ has been 
observed between $\sqrt{s_{NN}} = 11.5$ GeV and $\sqrt{s_{NN}} = 19.6$ GeV. This has been attributed to
the softening of the QCD equation of state (EoS)~\cite{Steinheimer:2022gqb,Nara:2016phs}. From transport model 
calculations, it has been observed that the rapidity odd directed flow is sensitive to the nature of EoS~\cite{Steinheimer:2022gqb,Nara:2016phs,Konchakovski:2014gda,Ivanov:2014ioa,Nara:2021fuu}
and provides a possible hint of first order phase transition of the QCD matter at larger baryon chemical potential~\cite{Steinheimer:2022gqb,Nara:2016phs}. It has been further noticed that the reported experimental 
data shows a splitting between the slope of baryon and anti-baryon directed flow which increases by decreasing 
the collision energy. Recently, a hydrodynamic model calculation at $\sNN=200$ GeV has shown that one may 
also obtain such splitting from the initial inhomogeneous deposition of baryon in the transverse plane~\cite{Bozek:2022svy}.

The directed flow is generated as a response to the initial pressure asymmetry and carries information about the early stage 
of collisions~\cite{Csernai:1999nf,Snellings:1999bt,Adil:2005qn,Becattini:2015ska,Bozek:2010bi,Ryu:2021lnx,Jiang:2021ajc}. 
Therefore, model-to-data comparison of the $v_1$ of identified hadrons has the potential to constrain the 
initial profile of both matter and baryon which in turn serves as a crucial ingredient to study the dynamics of conserved 
charges. 

There have been several attempts to capture the beam energy dependence of directed flow of identified particles in 
transport~\cite{Chen:2009xc,Guo:2012qi,Nara:2022kbb} and hybrid 
models~\cite{Steinheimer:2014pfa,Ivanov:2014ioa,Shen:2020jwv,https://indico.cern.ch/event/895086/contributions/4712143/}. 
However, a consistent approach that qualitatively captures the experimental trends across beam energies  is 
missing~\cite{Singha:2016mna}.

There have been efforts to construct frameworks for the three-dimensional evolution of the QCD conserved charges along with energy 
density for the RHIC BES
~\cite{Karpenko:2015xea,Du:2019obx,Li:2018fow,Shen:2020jwv,Denicol:2018wdp,Wu:2021fjf,De:2022yxq,Fotakis:2019nbq}. 
One of the important goals in building such frameworks is to constrain the transport coefficients related
to charge diffusion. Such attempts are hindered mainly by the large uncertainties in the initial matter and baryon deposition. 
An additional complication arises due to the large passage time of the colliding nuclei at lower $\sNN$, which consequently
increases the initial proper time when hydrodynamics becomes applicable. There have been few attempts to implement a 
dynamical initialization condition for hydrodynamic evolution~\cite{Akamatsu:2018olk,Du:2018mpf,Shen:2022oyg,De:2022yxq,Okai:2017ofp,Shen:2022oyg, Shen:2017bsr}. 
There have also been attempts to propose a simple geometrical ansatz that can qualitatively capture the trends in the data~\cite{Denicol:2018wdp,Shen:2020jwv,Ryu:2021lnx,Bozek:2022svy,https://indico.cern.ch/event/895086/contributions/4712143/}. 
In this work, we follow the latter approach. We have recently proposed an initial baryon profile~\cite{our_paper} which 
when coupled to tilted matter profile~\cite{Bozek:2010bi} provides a reasonably good description of the identified 
hadron $v_1$ at $\sNN=19.6$ GeV and 200 GeV. In this work, we have tested the performance of this newly proposed initial condition model 
across beam energies ranging from $\sqrt{s_{NN}} = $ 200 GeV to 7.7 GeV. In addition to that, the centrality and transverse momentum 
dependency of $v_1$ have been presented and compared to available measurements. 

In this study, we have used a multistage hydrid framework (hydrodynamic evolution + hadronic transport) for simulations at different $\sNN$. 
In the next section, we will describe the initial condition model which is used as an input to the hybrid framework. The transport coefficients and 
EoS used during the hydrodynamic evolution are presented in Sec.~\ref{sec3}. The procedure for selecting the model parameters to capture the experimental
results are explained in Sec.~\ref{sec4}. We have presented the results in Sec.~\ref{sec5} and Sec.~\ref{sec6} is devoted to summarize the current study with 
some concluding remarks. 

\section{Initial condition}
\label{sec2}
We have adopted a similar procedure as described in~\cite{Shen:2020jwv,Denicol:2018wdp} to prepare an event average profile
of the initial energy and the net baryon density. The participant and binary collision sources obtained from each MC Glauber event 
are rotated by the second-order participant plane angle and then smeared out in the transverse plane. The smearing profile 
is assumed to be a Gaussian with parametric width, $\sigma_{\perp}$. Profiles are prepared by averaging over 25,000 initial
configurations. 

Assuming asymmetric matter deposition by a participant along the rapidity, the form of the 
initial energy density $\epsilon(x,y,\eta_s; \tau_0)$ deposition at a constant proper time $\tau_0$ is 
taken as~\cite{Bozek:2010bi}, 
\beqa
  \epsilon(x,y,\eta_{s}) &=& \epsilon_{0} \left[ \left( N_{+}(x,y) f_{+}(\eta_{s}) + N_{-}(x,y) f_{-}(\eta_{s})  \right)\right.\nn\\
                           &&\left.\times \left( 1- \alpha \right) + N_{coll} (x,y)  \epsilon_{\eta_s}\left(\eta_{s}\right) \alpha \right] 
 \label{eq.tilt}
\eeqa
where, $N_{+}(x,y)$  and $N_{-}(x,y)$ are the participant densities of forward and 
backward moving nucleus respectively. $N_{coll} (x,y)$  is the contribution
of binary collision sources at each transverse position $(x,y)$. $\alpha$ 
is the hardness factor. $ f_{+,-}(\eta_s)$ are the asymmetric rapidity 
envelop function for the energy density, $\epsilon$. 
\begin{equation}
    f_{+,-}(\eta_s) = \epsilon_{\eta_s}(\eta_s) \epsilon_{F,B}(\eta_s)
\end{equation}
with
\begin{equation}
    \epsilon_{F}(\eta_s) = 
    \begin{cases}
    0, & \text{if } \eta_{s} < -\eta_{m}\\
    \frac{\eta_{s} + \eta_{m }}{2 \eta_{m}},  & \text{if }  -\eta_{m} \le \eta_{s} \le \eta_{m} \\
    1,& \text{if }  \eta_{m} < \eta_{s}
\end{cases}
\end{equation}
and 
\begin{equation}
    \epsilon_{B} (\eta_s) = \epsilon_F(-\eta_s)
\end{equation}

The form of the initial baryon profile is
\begin{equation}
    n_{B} \left( x, y, \eta_s \right) = 
       N_{B} \left[ W_{+}^{B}(x,y) f_{+}^{B}(\eta_{s}) + W_{-}^{B}(x,y) f_{-}^{B}(\eta_{s})  \right]
    \label{my_baryon_ansatz}
\end{equation}
$W_{\pm}^{B}(x,y)$ are the weight factors to deposit the net baryon in the transverse plane which has the following form.

\begin{equation}
W_{\pm}^{B}(x,y) = \left( 1 - \omega \right) N_{\pm}(x,y) + \omega N_{coll}(x,y)
    \label{weight_ansatz_1_for_baryon}
\end{equation}
The net baryon distribution in the transverse plane can be tuned by varying the phenomenological parameter $\omega$. 
The net baryon density rapidity envelope profiles are taken as~\cite{Denicol:2018wdp,Shen:2020jwv}, 
\beqa
    f_{+}^{n_{B}} \left( \eta_s \right) &=&  \left[  \theta\left( \eta_s - \eta_{0}^{n_{B} } \right)   \exp{- \frac{\left( \eta_s - \eta_{0}^{n_{B} }  \right)^2}{2 \sigma_{B, + }^2}}   + \right.\nn\\ && \left. \theta\left(  \eta_{0}^{n_{B} } - \eta_s \right)   \exp{- \frac{\left( \eta_s - \eta_{0}^{n_{B} }  \right)^2}{2 \sigma_{B, - }^2}}   \right]
\label{forward_baryon_envelop}
\eeqa
and
\beqa
    f_{-}^{n_{B}} \left( \eta_s \right) &=&  \left[   \theta\left( \eta_s + \eta_{0}^{n_{B} } \right)   \exp{- \frac{\left( \eta_s + \eta_{0}^{n_{B} }  \right)^2}{2 \sigma_{B, - }^2}}   + \right.\nn\\ && \left. \theta\left( -\eta_s -  \eta_{0}^{n_{B} }  \right)   \exp{- \frac{\left( \eta_s + \eta_{0}^{n_{B} }  \right)^2}{2 \sigma_{B, + }^2}}   \right]
\label{backward_baryon_envelop}
\eeqa

The normalization factor $N_B$ in Eq.~\ref{my_baryon_ansatz} is not a free parameter, rather 
it is constrained by the initially deposited net baryon carried by the participants. 
\begin{equation}
      \int  \tau_{0}  n_{B} \left( x, y, \eta, \tau_{0} \right) dx  dy  d\eta  = N_{\text{Part}}
      \label{net_baryon_is_npart_2}
\end{equation}
With the asymmetric baryon profile given in Eqs.~\ref{forward_baryon_envelop} and~\ref{backward_baryon_envelop} we 
generate a tilted baryon profile in the reaction plane at the initial stage. The magnitude of the tilt can be controlled  
by changing the $\omega$ parameter. 

\section{Hydrodynamic Evolution}
\label{sec3}
The publicly available MUSIC~\cite{Schenke:2010nt, Paquet:2015lta, Schenke:2011bn} code has been used for 
the hydrodynamic evolution of the deposited energy and the baryon density profile. The hydrodynamic equations of
the conserved quantities and the dissipative currents in the presence of finite baryon density, which are solved 
in the framework of MUSIC have been described in~\cite{Denicol:2018wdp}. Other conserved charges, net 
strangeness( $n_S$) and net electric charge densities($n_Q$) are not evolved independently and 
are assumed to satisfy the following constraints locally. 
\beqa
n_S&=&0\label{eq.ns}\\ 
n_Q&=&0.4n_B\label{eq.nq}
\eeqa

A temperature ($T$) and baryon chemical potential ($\mu_B$) dependent baryon diffusion coefficient($\kappa_{B}$) 
has been implemented in the code which is derived from the Bolzmann equation in relaxation time approximation~\cite{Denicol:2018wdp}.
The form of $\kappa_B$ is as follows :
\beq
\kappa_{B} = \frac{C_B}{T} n_{B} \left[ \frac{1}{3} \coth{\left(\frac{\mu_B}{T}\right)} - \frac{n_B T}{\epsilon + p} \right]
\eeq
$C_B$ is a free parameter which controls the baryon diffusion in the medium. This has been taken as a model parameter in the 
simulation. In the above expression, $n_B$ is the net baryon density and $p$ is the local pressure of the fluid. The 
specific shear viscosity $C_{\eta}$ is related to the shear transport co-efficient $\eta$ as follows.
\beq
C_{\eta} = \frac{ \eta T}{\epsilon + p}
\eeq 
This is another model parameter and chosen to be 0.08. In this work, we have not considered the effects of bulk viscosity. 
%

A lattice QCD based EoS at finite baryon density, NEoS-BQS~\cite{Monnai:2019hkn, HotQCD:2012fhj, Ding:2015fca, Bazavov:2017dus} 
has been used during the hydrodynamic evolution. The EoS imposes the constraints of Eqs.~\ref{eq.ns} and \ref{eq.nq}

The Cooper-Frye conversion of fluid into particles has been performed on the hypersurface of constant energy density, 
$\epsilon_{f} = 0.26$ GeV/fm$^{3}$ using iSS~\cite{https://doi.org/10.48550/arxiv.1409.8164,https://github.com/chunshen1987/iSS}. 
The sampled primary hadrons are then fed into UrQMD~\cite{Bass:1998ca, Bleicher:1999xi} for the late stage hadronic 
transport.   

\section{Model parameters}
\label{sec4}
We have studied Au+Au collisions at seven different energies: 7.7, 11.5, 19.6, 27, 39, 62.4 and 200 GeV. 
The values of the model parameters used in the study have been summarized 
in Table~\ref{param_for_model}. They are chosen to describe the experimentally observed
rapidity dependence of the charged particle yield, net-proton yield and directed flow of $\pi^{+}, p$ and $\bar{p}$ 
simultaneously.  Using these model parameters, we have computed and presented the rapidity dependence of $v_1$ of 
other identified hadrons: K$^\pm$, $\Lambda$, $\bar{\Lambda}$ and $\phi$. In addition, we have also presented the $p_T$ 
differential $v_1$.

In our previous work~\cite{our_paper}, we have shown that our proposed initial condition with both zero 
and non-zero $C_B$ is able to describe the observed splitting between baryon and anti-baryon directed flow 
at $\sqrt{s_{NN}} = $19.6 GeV and 200 GeV within the experimentally measured rapidity range. 
Hence, we are not able to constrain the $C_B$ values within the current 
ambit of model and available experimental data by a unique model-to-data comparison.
The same has been observed for other $\sNN$. 
In this work, we have presented the model calculations with $C_B=1$ for $\sNN$ above 11.5 GeV and $C_B=0.5$ at 
$\sNN=11.5$ and 7.7 GeV (we couldn't find a suitable parameter set for $C_B=1$ at these lowest energies).

The normalization parameter of the initial energy density distribution $\epsilon_{0}$ is calibrated to 
match the yield of charged hadron in mid-rapidity. However, the $\pi^{+}$ yield has been considered to set $\epsilon_0$ at
energies where the mid-rapidity measurement of charged hadron is not available. The hardness factor $\alpha$ has been 
chosen to capture the centrality dependence of the charged hadron yield. 

The plateau length $\eta_{0}$ and Gaussian fall off $\sigma_{\eta}$ in the initial 
matter profile are adjusted to describe the rapidity dependence of the charged particle yield. 
From the existing experimental data, we have observed that the rapidity dependent charged hadron yields 
at different energies follow the same distribution when the pseudo-rapidity is scaled by the 
respective beam rapidity. Therefore, we have chosen $\eta_{0}$ and $\sigma_{\eta}$ to capture that scaled 
distribution at the energies where the rapidity distribution of neither the charged particle 
nor the identified particle has been experimentally measured. 

Rapidity distribution of net-proton yield which is capable of constraining
 $\eta_{0}^{n_{B}}$, $\sigma_{B,+}$ and $\sigma_{B,-}$ has been measured 
for Au+Au systems at $\sqrt{s_{NN}} = 200 $ and 62.4 GeV by the BRAHMS 
collaboration~\cite{BRAHMS:2003wwg,BRAHMS:2009wlg} but is not available
at other energies considered in this work. However, the mid-rapidity 
measurements for Au+Au systems has been done by STAR collaboration~\cite{STAR:2017sal}.
We tuned the parameters of initial baryon profile to match with the mid-rapidity measurements at 
energies below 62.4 GeV.

The adopted initial condition model creates a tilted profile of both energy density and the
net baryon density in the reaction plane (spanned by beam axis and impact parameter axis).
Individually $\eta_m$ and $\omega$ parameters can be tuned to obtain the desired tilt
in the energy and baryon profile respectively. However, only a proper set of $\eta_m$ and $\omega$ 
choice can explain the directed flow of $\pi^{+},p$ and $\bar{p}$ simultaneously after hydrodynamic evolution.

\begin{table}[h]
\begin{tabular}{|p{0.9cm}|p{0.4cm}|p{0.7cm}|p{1.1cm}|p{0.55cm}|p{0.4cm}|p{0.4cm}|p{0.5cm}|p{0.6cm}|p{0.6cm}|p{0.45cm}|p{0.55cm}|}
\hline 
$\sqrt{S_{NN}}$ \tiny{(GeV)} & $C_B$ & $\tau_0$\tiny{(fm)} &$\epsilon_{0}$ \tiny{(GeV/fm$^{3}$)} & $\alpha$ &  $\eta_{0}$ & $\sigma_{\eta}$ & $\eta_{0}^{n_{B}}$ & $\sigma_{B,-}$ & $\sigma_{B,+}$ & $\eta_m$ & $\omega$ \\ \hline
200  & 1.0 & 0.6  &  8.0 & 0.14  &  1.3  &  1.5  &  4.6  &  1.6   &  0.1  & 2.2 & 0.25  \\ 
\hline
62.4 & 1.0 & 0.6  &  5.4 & 0.14  &  1.4  &  1.0  &  3.0  &  1.0   &  0.1  & 1.4 & 0.25  \\
\hline 
39   & 1.0 & 1.0  &  3.0 & 0.12   &  1.0  &  1.0  &  2.2  &  1.2   &  0.2  & 1.1 & 0.2  \\ 
\hline
27   & 1.0 & 1.2  &  2.4 & 0.11  &  1.3  &  0.7  &  2.3  &  1.1   &  0.2  & 1.1 & 0.11  \\ 
\hline
19.6 & 1.0 & 1.8  &  1.55 & 0.1  &  1.3  &  0.4  &  0.3  &  0.8   &  0.15 & 0.8 & 0.15  \\ 
\hline
11.5 & 0.5 & 2.6  &  0.9 & 0.1  &  0.9  &  0.4  &  1.2  &  0.55  &  0.2  & 0.4 & 0.22   \\
\hline 
7.7  & 0.5 & 3.6  &  0.55 & 0.1  &  0.9  &  0.4  &  0.9  &  0.35  &  0.2  & 0.3 & 0.35  \\ 
\hline
\end{tabular}
\caption{ Model parameters used in the simulations at different $\sqrt{S_{NN}}$ . }
\label{param_for_model}
\end{table}

\section{Results}
\label{sec5}
\begin{figure*}
 \begin{center}
 \includegraphics[scale=0.56]{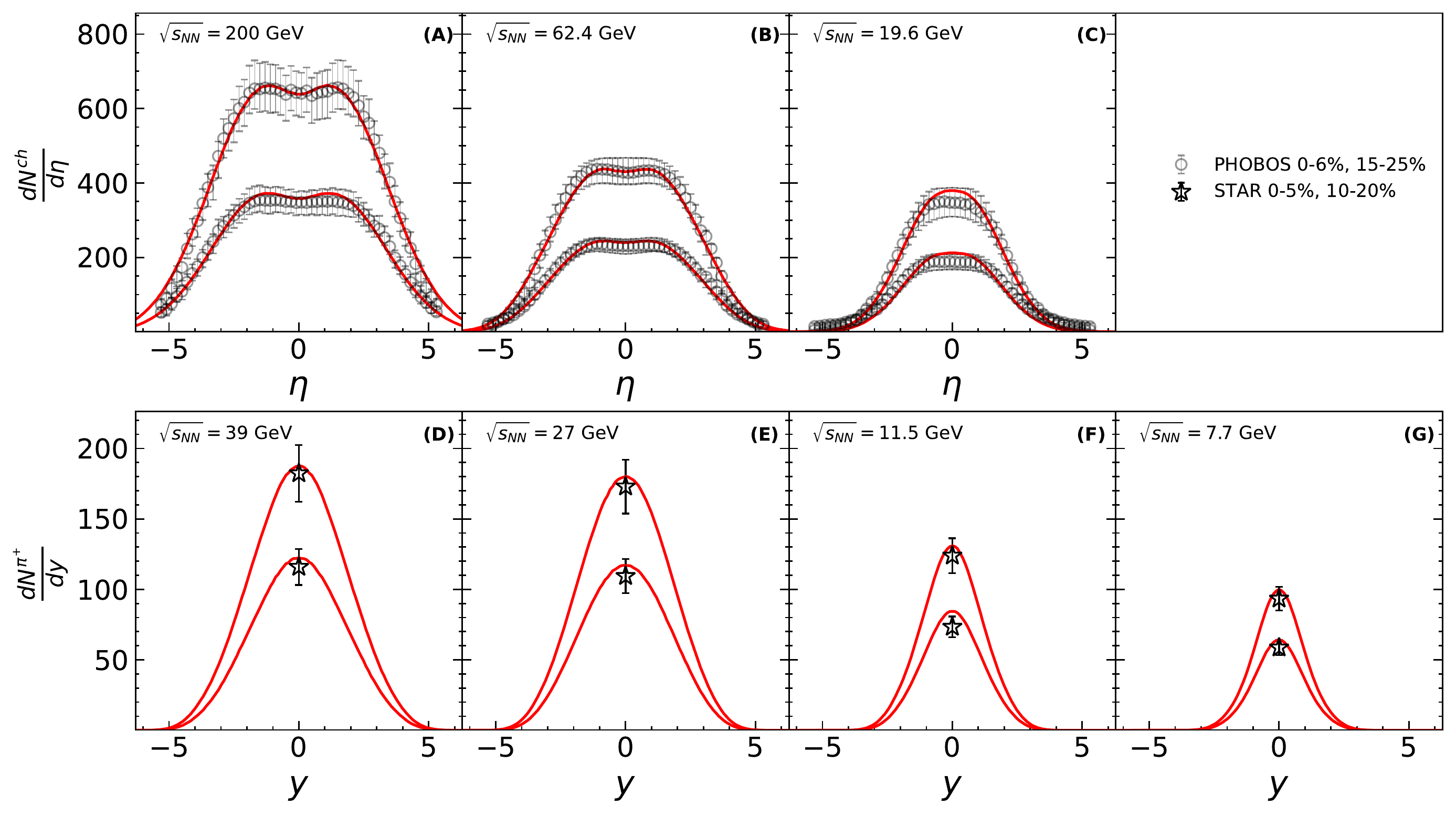}
 \caption{(Color online)  Pseudo-rapidity distribution of charged hadrons in 0-6$\%$, 15-25$\%$ Au+Au collisions at $\sNN=200, 62.4$ and $19.6$ GeV has been shown in panel (A), (B) and (C) respectively. The model calculations(red solid lines) are compared with the measurements from PHOBOS collaboration~\cite{Back:2002wb}. The $\pi^{+}$ yield has been plotted a function of rapidity for $\sNN=39, 27, 11.5$ and $7.7$ GeV in panel (D),(E),(F) and (G). Comparisons are made with the mid-rapidity mesurements by STAR collaboration~\cite{STAR:2017sal}. 
     }
 \label{fig1_dnchdeta_eta}
 \end{center}
\end{figure*}

Pseudo-rapidity($\eta$) distributions of charged hadrons have been plotted in panel (A), (B) and (C) of Fig.~\ref{fig1_dnchdeta_eta}
for 0-6$\%$ and 15-25$\%$ Au+Au collisions at $\sqrt{s_{NN}} = 200, 62.4$ and $19.6$ GeV. We are able to capture the rapidity distribution
by suitably choosing the $\epsilon_{0}$, $\eta_{0}$ and $\sigma_{\eta}$ for the initial space-time rapidity distribution of energy density. 
The rapidity($y$) differential $\pi^{+}$ yield has been presented in panel (D), (E), (F) and (G) of Fig.~\ref{fig1_dnchdeta_eta} for Au+Au 
collisions at $\sqrt{s_{NN}} = 39, 27, 11.5 $ and $7.7$ GeV where the charged particle measurement is not available. We have fixed the 
parameter $\epsilon_{0}$ by matching the model calculation with the mid-rapidity $\pi^{+}$ yield measurement at those energies. The rapidity
distribution of inital energy profile has been calibrated by compairing the model calculation with the scaled distribution as discussed in
previous section. The chosen hardness factor $\alpha$ provides a reasonable description about the centrality dependence of the charged particle 
yield across all energies.

\begin{figure*}
 \centering
 \includegraphics[width=0.87\textwidth]{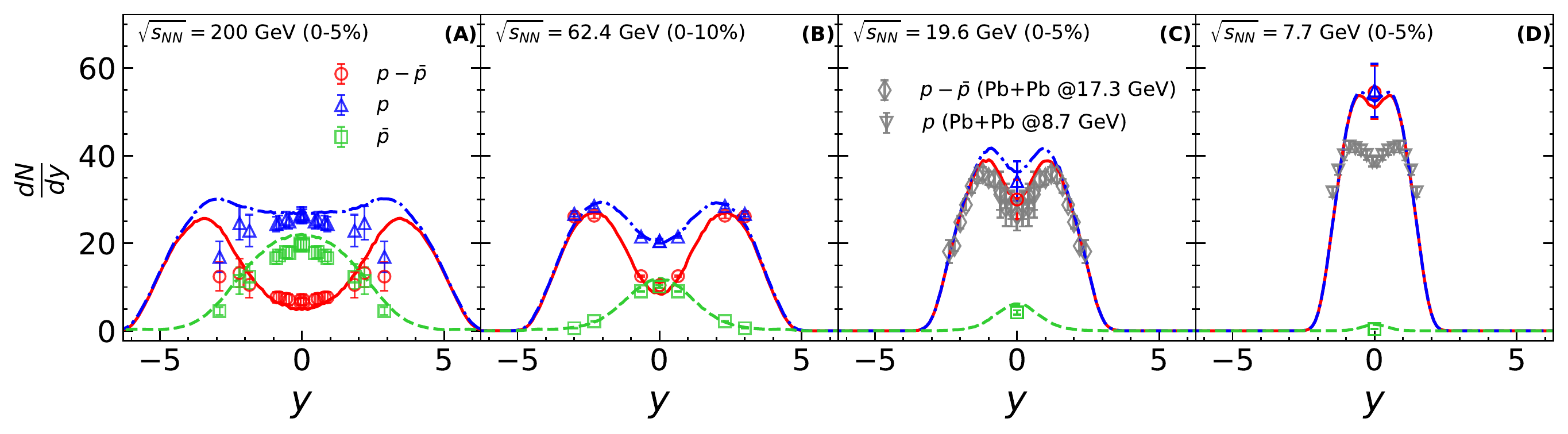}
 \caption{(Color online) Rapidity distribution of proton, anti-proton and net-proton for 0-5$\%$ Au+Au collisions at $\sqrt{s_{NN}} = 200, 19.6, 7.7$ GeV and for 0-10$\%$ Au+Au colllisions at $\sqrt{s_{NN}} = 62.4$ GeV. The net-proton rapidity distributions at $\sqrt{s_{NN}} = 17.3$ and $8.7$ GeV for Pb+Pb systems are plotted in panel (C) and (D) respectively for reference. The measurements are from~\cite{BRAHMS:2003wwg, BRAHMS:2009wlg, STAR:2017sal, NA49:2010lhg}. The model calculations of proton(blue dashed-dotted line), anti-proton(green dashed line) and net-proton(red solid line) are compared with the experimental measurements.}
 \label{fig3_p_pbar_netp_dy}
\end{figure*}

It is important to look at the distribution of net-proton along rapidity to 
constrain the longitudinal gradient of baryon chemical potential which plays a significant role in flow calculations. In this regard, we have plotted
the proton, anti-proton and net-proton rapidity distribution for central Au+Au collsions at $\sqrt{s_{NN}} = 200, 62.4$, $19.6$ and $7.7$ GeV in Fig.~\ref{fig3_p_pbar_netp_dy}. We have obtained a good agreement between our model calculations and the experimental measuremets. The contributions from weak decays have been included in the calculation of proton and anti-proton yield to compare with the experimental measurements. 
In addition to that, we are able to capture the rapidity distribution of proton and anti-proton separately which reflects the fact that the chosen freeze-out
energy density represents a proper combination of the tempearture and baryon chemical potential of chemical equilibration~\cite{Andronic:2017pug}. The rapidity diferential net-proton measurements for Pb+Pb collisions at $\sqrt{s_{NN}}=$ 17.3 GeV and 8.7 GeV has been done by NA49 collaborartion~\cite{NA49:2010lhg}. We have considered the experimental data of net-proton distribution at 17.3 GeV as a proxy to constrain the model parameters present in Eq.~\ref{forward_baryon_envelop} and ~\ref{backward_baryon_envelop} for the Au+Au collisions at 19.6 GeV, whereas the data of net-proton at 8.7 GeV has been put along with the model 
calculation of Au+Au $\sqrt{s_{NN}} = 7.7$ GeV just for reference.

\begin{figure*}
 \centering
 \includegraphics[width=1.0\textwidth]{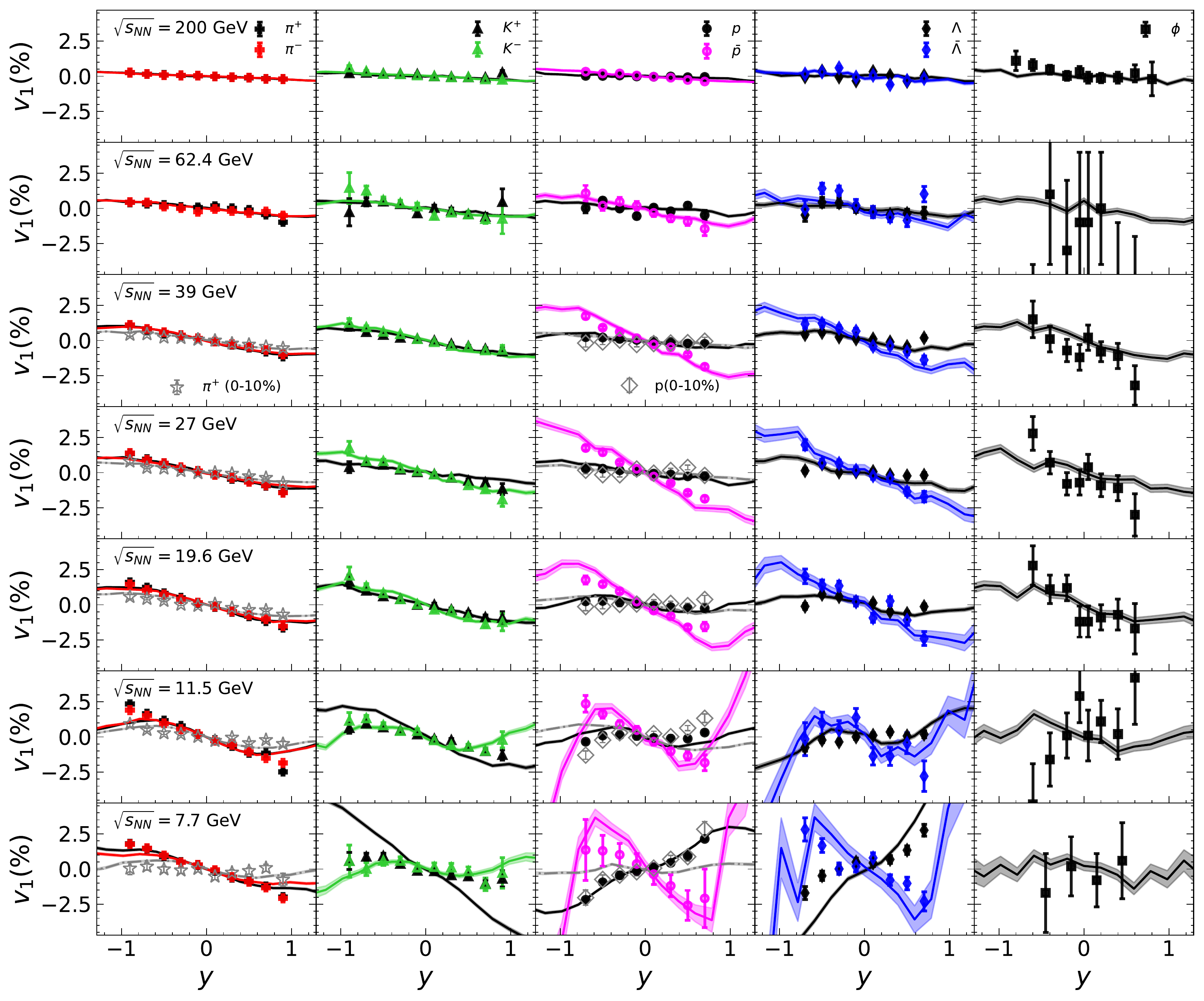}
 \caption{(Color online) Rapidity dependence of identified particles' directed flow coefficient($v_1$) for 10-40$\%$ Au+Au collisions at $\sqrt{s_{NN}} = 200, 62.4, 39, 27, 19.6, 11.5$  and $7.7$ GeV. Plots for a particular energy are placed in a single row of the figure. Model calculations(lines with shaded bands) are compared with the experimental measurements(different symbols) of STAR collaboration~\cite{STAR:2014clz,STAR:2017okv}. The available measurements and model calculations for $0-10\%$ centrality are plotted in grey colored symbols and lines.}
 \label{fig4_v1_y}
\end{figure*}

After calibrating the model parametrs of the initial energy and baryon density profile, we now present the directed flow($v_1$) of 
identified particles in 10-40$\%$ Au+Au collisions at different $\sqrt{s_{NN}}$ in Fig.~\ref{fig4_v1_y}. Each row in the figure contains the 
results from a particluar collision energy whereas the directed flow of various particle species are put in different columns.
The top row contains the rapidity dependence of directed flow coefficients in Au+Au collisions at $\sqrt{s_{NN}}=200$ GeV. 
The results for other energies are placed in subsequent rows in descending order. The $v_1$ for $\pi^{\pm}$, 
$K^{\pm}$, $p-\bar{p}$, $\Lambda-\bar{\Lambda}$ and $\phi$ has been presented in column 1, 2, 3, 4 and 5 respectively. In addition, the 
$v_{1}$ of $\pi$ and $p$ in 0-10$\%$ centrality class has been plotted for the comparison at energies below $\sqrt{s_{NN}} = 39$ GeV.

The relative tilt between the matter and baryon profiles determines the sign of the splitting between proton and anti-proton directed
flow~\cite{our_paper}. Thus, by suitable choice of ($\eta_{m}$, $\omega$) at each $\sqrt{s_{NN}}$, we are able to describe the rapidity 
dependence of $v_1$ for $\pi^{+}, p $ and $\bar{p}$ simultaneously in 10-40$\%$ centrality whereas, the $v_1$ of other particle species 
plotted in Fig~\ref{fig4_v1_y} are the model predictions. We are able to capture the sign change in slope of proton directed flow at 
$\sqrt{s_{NN}} = 7.7$ GeV. From Table~\ref{param_for_model}, one can observe that the value of $\eta_m$ decreases consistently 
with collision energy whereas the chosen $\omega$ value for model calibration seems to suggest a non-monotonic behaviour with 
collision energy. In order to confirm such non-monotonic behaviour of $\omega$ with $\sNN$, we need to perform a more sophisticated 
fitting procedure to extract the model parameters. The purpose of the current study is to demonstrate the existence of reasonable 
parameter space in this proposed model that captures the data trends quite well.


\begin{figure}
 \begin{center}
 \includegraphics[scale=0.4]{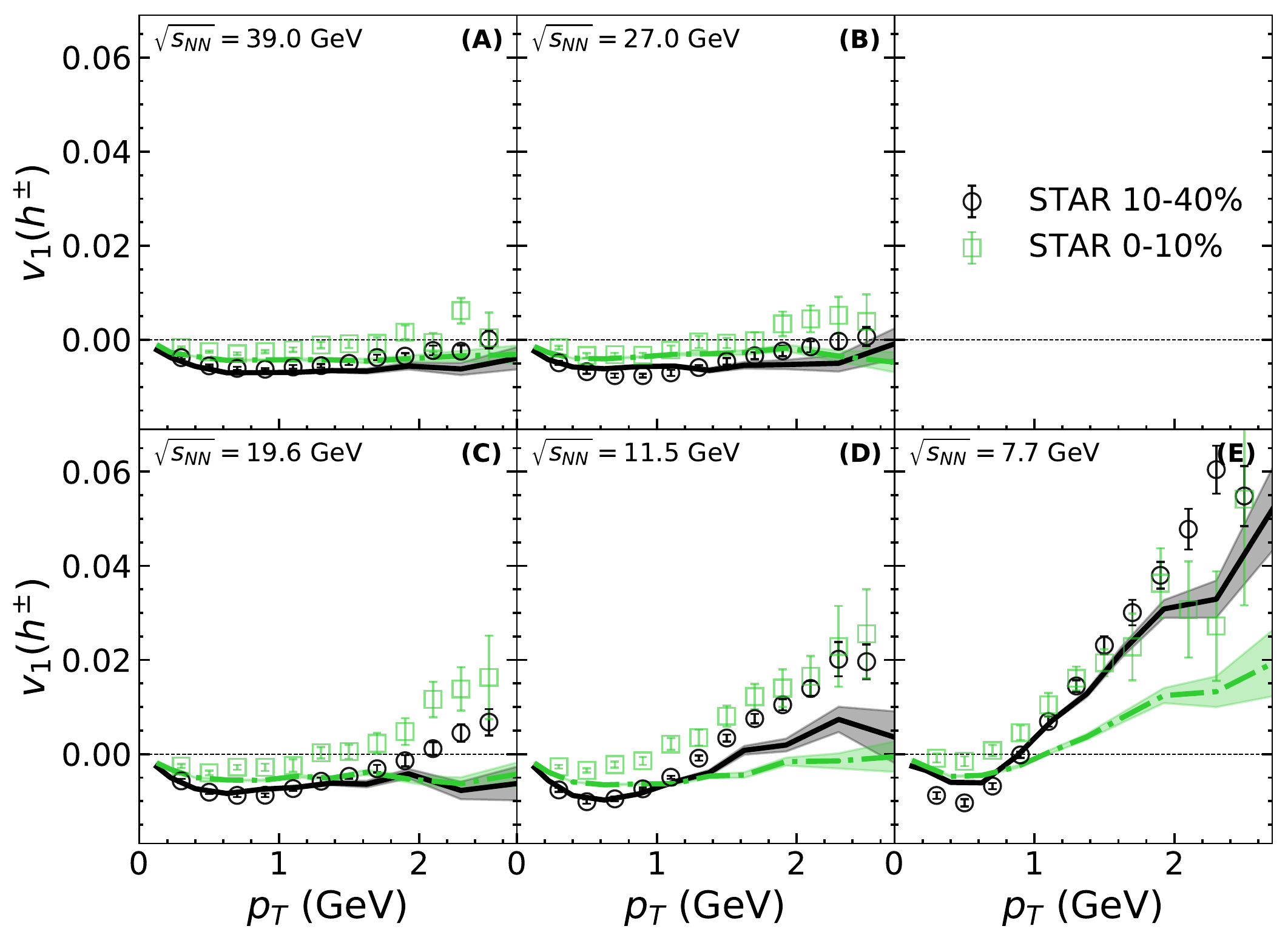}
 \caption{(Color online) Transverse momentum dependency($p_T$) of charged particle directed flow for 0-10$\%$ and 10-40$\%$ Au+Au collisions at $\sqrt{s_{NN}}=39, 27, 19.6, 11.5$  and $7.7$ GeV. The model calculations for 0-10$\%$(green dashed-dotted lines) and 10-40$\%$(black solid lines) centrality are compared with measurements from STAR collaboration~\cite{STAR:2019vcp}.    }
 \label{v1_ch_pT}
 \end{center}
\end{figure}

\begin{figure}
 \begin{center}
 \includegraphics[scale=0.22]{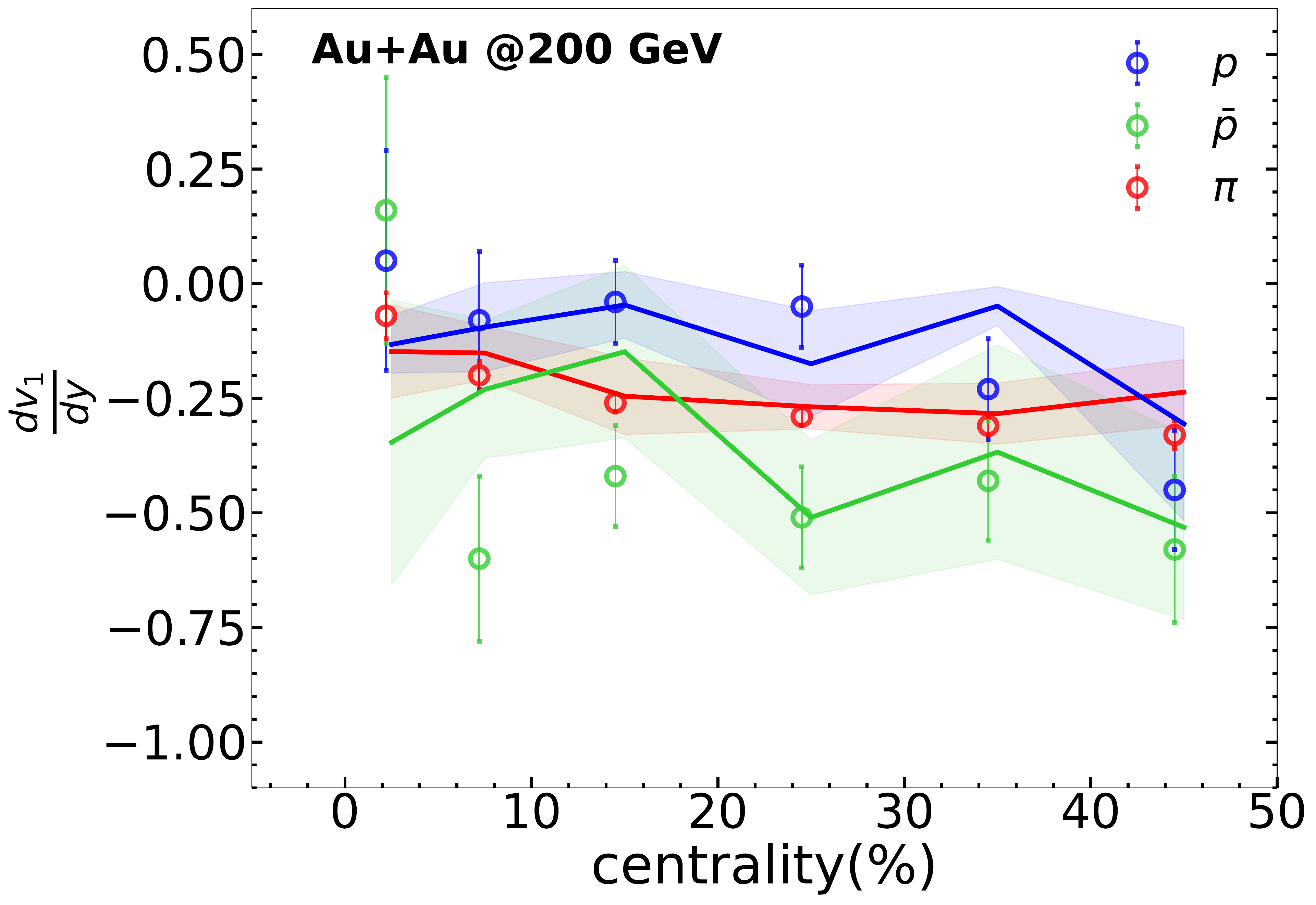}
 \caption{(Color online) Centrality dependency of identified particles' directed flow coefficient($v_1$) for Au+Au collisions at $\sqrt{s_{NN}} = 200$ GeV. Model calculations of $\pi^{+}$(red line), $p$(green line) and $\bar{p}$(blue line) are compared with experimental data from~\cite{STAR:2011hyh}. }
 \label{dv1dy_cent_200GeV}
 \end{center}
\end{figure}

An inhomogeneous deposition and further evolution of net-baryon density in the medium creates inhomogeneities for the other conserved charges 
like strangeness and electric charge via the constraints in Eqs.~\ref{eq.ns} and ~\ref{eq.nq} that give rise to correlations between the 
corresponding chemical potentials: $\mu_B$, $\mu_Q$ and $\mu_S$ for baryon, electric charge and strangeness respectively. This gives rise to 
splitting in directed flow of hadrons with different quantum numbers but same mass similar to the $v_1$ splitting of $p$ and $\bar{p}$.
The splitting of $v_1$ between $\pi^{+}$ and $\pi^{-}$ is solely due to the inhomogeneity of $\mu_{Q}$ in the medium. However, the inhomogeneity of both $\mu_S$ and $\mu_Q$ are responsible for the difference in directed flow coefficient of $K^{+}$ and $K^{-}$. In the present model calculation, we have not observed any significant splitting of $v_1$ between $\pi^{+}$ and $\pi^{-}$ which is consistent with experimental measurements. However, the splitting between $K^{+}$ and $K^{-}$ at $\sqrt{s_{NN}} = 11.5 $ and $7.7$ GeV is noticeable while it is consistent with zero in data. The same has been observed in case of $\Lambda$ and $\bar{\Lambda}$ since they are getting affected by $\mu_S$ along with $\mu_B$. In this case, the data also shows split. Although 
the model split is consistent with data for $\sNN>11.5$ GeV, for $\sqrt{s_{NN}} = 11.5 $ and $7.7$ GeV the model overpredicts the data. At these same 
energies, the model also fails to capture the $\phi$ $v_1$ implying interesting physics of strange carriers that has not been included in the 
current model. These discrepancies underline the significance to evolve all the conserved charges independently in a fluid dynamical simulation~\cite{Greif:2017byw,Fotakis:2019nbq}.  

The directed flow measurements of $\pi$ and $p$ at 0-10$\%$ centrality are also plotted in Fig.~\ref{fig4_v1_y} along with 10-40$\%$ for the ease of comparison. Our model calculations capture the centrality dependence of $v_1$ for $\pi$ at all the considered collision energies but it fails to do so for $p$ below $\sqrt{s_{NN}} = 19.6$ GeV. It indicates that the mechanism of baryon stopping from central to peripheral collisions is different at lower energies and our model is unable to provide a proper gradient of initial net-baryon density at different centralities. From the microscopic models, we have the guidance that the baryon deposition mechanism is dependent on transverse position~\cite{Li:2018ini,Li:2016wzh,De:2022yxq,Shen:2017bsr,Shen:2022oyg} which need to be further explored. In this direction the centrality dependence measurement of proton directed flow could play a significant role.

So far we have discussed about the rapidity differential behaviour of $v_1$. Now we will present the $p_T$ dependency of $v_1$ which provides the information about the initial deposition and evolution of matter in the transverse plane around mid-rapidity region. In this regard, The $p_{T}$ differential directed flow of charged hadrons in 0-10$\%$ and 10-40$\%$ Au+Au collisions at $\sqrt{s_{NN}}=39, 27, 19.6, 11.5$ and $7.7$ have been plotted
in Fig.~\ref{v1_ch_pT}. The model calculations are in agreement with the experimental measurements in low $p_T$ region at all energies but fail to explain the trend
in the high $p_T$ region. Nevertheless, the $p_T$ integrated $v_{1}$ does not get affected by this discrepancies due to relatively higher yield of low $p_T$ hadrons. We are able to capture the centrality trend in the $p_T$ differential $v_1$, which in turn shows the correct centrality dependence of rapidity differential $\pi$ directed flow in Fig~\ref{fig4_v1_y}.

Fig.~\ref{dv1dy_cent_200GeV} shows the centrality dependence of the slope of identified particle directed flow in Au+Au collisons at $\sqrt{s_{NN}}=200$ GeV. $v_1$ slopes in the model at each centrality have been calculated by fitting a straight line in the mid-rapidity region. Although the model parameter is tuned for 10-40$\%$ centrality, we are able to describe the magnitude and hierarchy in the directed flow slope of $\pi, p $ and $\bar{p}$ at other centralities.

\begin{figure}
 \begin{center}
 \includegraphics[scale=0.4]{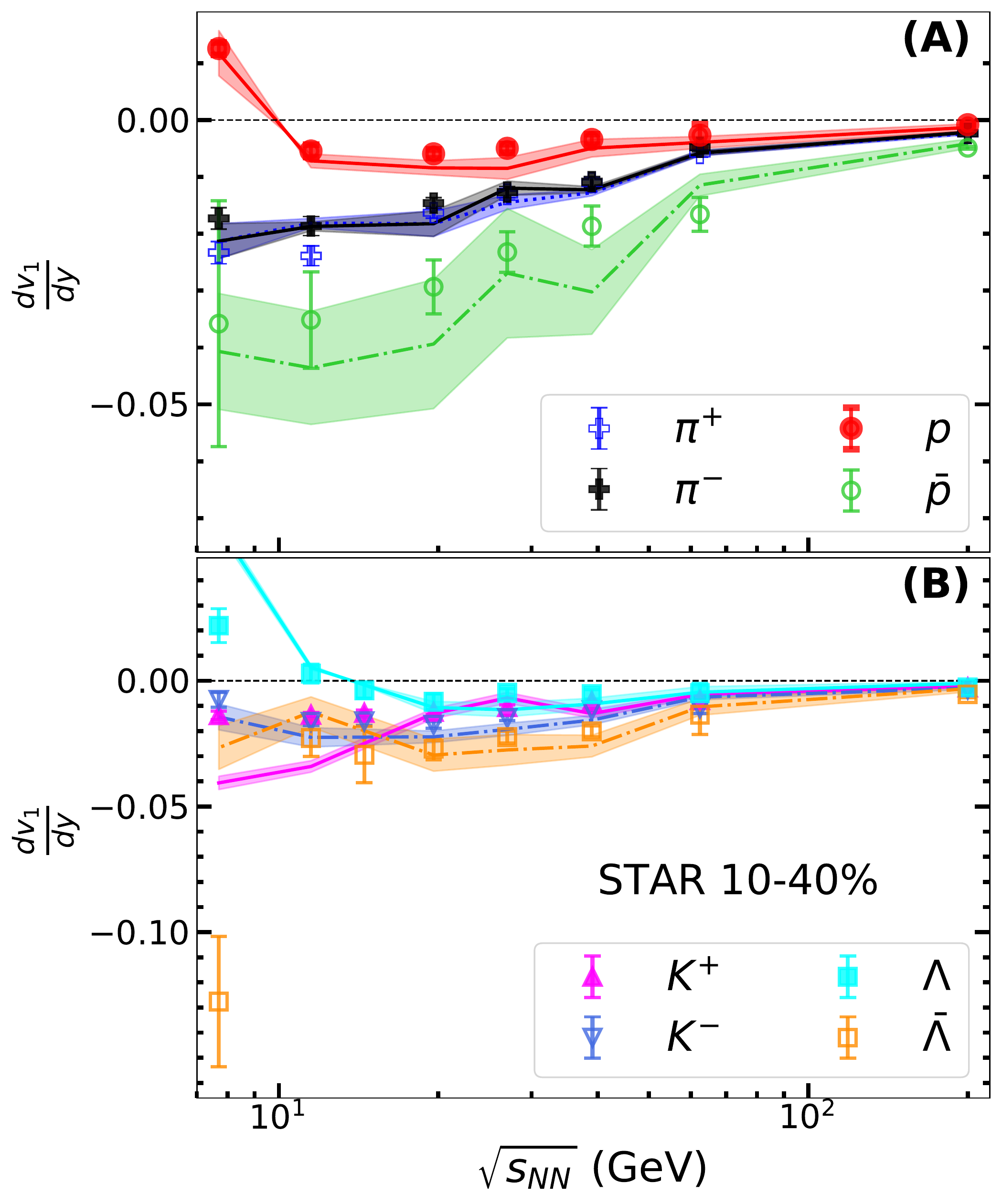}
 \caption{(Color online) Beam energy dependence of identified particles' directed flow slope($\frac{dv_1}{dy}$) for 10-40$\%$ Au+Au collisions. The model calculation for a particular particle species is plotted as a line having the same color as the symbol of experimental data. The experimental meaurements are from STAR collaboration~\cite{STAR:2014clz,STAR:2017okv}.    }
 \label{dv1dy_pi_p_pbar_BES}
 \end{center}
\end{figure}

The slope of identified particles' directed flow has been plotted as function of collision energy for 10-40$\%$ Au+Au collisions in Fig.~\ref{dv1dy_pi_p_pbar_BES}.
Slopes of $\pi^{\pm}, p$ and $\bar{p}$ have been plotted in panel (A) whereas the slopes of $K^{\pm}, \Lambda$ and $\bar{\Lambda}$ have been presented in the panel (B). In the model calculation of directed flow slope, the fitting function and fitting range in rapidity has been taken the same as mentioned in the experimental papers~\cite{STAR:2014clz,STAR:2017okv}. We have observed that the $v_1$-slope of $p$ and $\Lambda$ changes sign at lower energies whereas the slope of their corresponding anti-particles remains negative at all energies. The mangnitude of the slopes are in good agreement with the expermental data. The split between the slopes of $K^{+}-K^{-}$ and $\Lambda-\bar{\Lambda}$ at $\sqrt{s_{NN}} = 7.7$ GeV is overestimed by the model calculations which is already observed in the plots of rapidity differential $v_1$ in Fig~\ref{fig4_v1_y}.

\section{Summary}
\label{sec6}
In the present study, our goal is to study the measured beam energy dependence of the rapidity-odd directed flow of identified particles at RHIC BES. In this regard, we have adopted the model of initial baryon deposition proposed in~\cite{our_paper}. The model assumes the asymmetric matter and baryon deposition in forward and backward rapidity by a participating nucleon.
As a consequence, a tilted matter and baryon profile in the reaction plane are formed, although the tilts are different. Taking the model as an input for a multi stage hybrid framework(hydrodynamics + hadronic transport) and calibrating the model parameters to describe the measured rapidity distribution of charged particle and net-proton,  we are able to describe the observed directed flow splitting between proton and anti-proton across beam energies ranging from 7.7 GeV to 200 GeV. The model is able to capture the sign change of proton-$v_1$ slope at lower enrgies. In addition to that we have studied the directed flow of $\Lambda, \bar{\Lambda}$, $\pi^{\pm}, K^{\pm}$ and $\phi$. From model calculations, we are able to describe the rapidity differential directed flow of all the identified particles simultaneously above $\sqrt{s_{NN}}=11.5$ GeV. However, the noticable discrepancies in the split of direct flow between $K^{+}-K^{-}$ and $\Lambda-\bar{\Lambda}$ at $\sqrt{s_{NN}}=7.7$ GeV is attributed to the constraints put in the EoS. This indicates that individual evolution of strangeness and electric charge should be done in hydrodynamic simulations at lower RHIC energies instead of constraing those via EoS. Our model is unable to capture the centrality dependency of proton directed flow at lower beam energies which shows that baryon stopping mechanism is very different from central to peripheral collisions. The analysis demonstrates the importance of the measurement of centrality dependecy of proton-$v_1$. 

In hydrodynamic calculations, it is known that the flow coefficients are response of initial gradient. Hence, the initial deposition plays a dominant role in explaining the measured directed flow of identified particles. Nevertheless, the effects due to non-trivial phenomena like dynamical initialization, non-zero initial longitudianl flow, event-by-event fluctuations and the nature of EoS which are not considered in present framework should not be ignored while studying the dynamics of conserved charges.

\bibliographystyle{apsrev4-1}
\bibliography{BES}

\end{document}